\documentclass[1p]{elsarticle}

\usepackage{hyperref}
\usepackage[T1]{fontenc} 
\usepackage{graphicx}
\usepackage{float}
\usepackage{adjustbox}
\usepackage{natbib}
\usepackage{hyperref}
\usepackage{amsmath}
\usepackage{amsthm}
\usepackage{mathtools}
\usepackage{adjustbox}
\usepackage{xcolor}
\usepackage{fancyhdr}
\allowdisplaybreaks

\biboptions{authoryear}

\newsavebox\MBox
\newcommand\Cline[2][red]{{\sbox\MBox{$#2$}%
  \rlap{\usebox\MBox}\color{#1}\rule[-1.2\dp\MBox]{\wd\MBox}{0.5pt}}}

\begin{document}

\begin{frontmatter}

\title{\href{https://doi.org/10.1016/j.radmeas.2018.07.016}{$\mu$Dose: a compact system for environmental radioactivity and dose rate measurement}}

\author[SUT]{Konrad Tudyka}
\ead{konrad.tudyka@polsl.pl}

\author[SUT]{Sebastian Mi\l{}osz}

\author[SUT]{Grzegorz Adamiec}

\author[SUT]{Andrzej Bluszcz}

\author[SUT]{Grzegorz Por\k{e}ba}

\author[AKOTECH]{\L{}ukasz Paszkowski}

\author[AKOTECH]{Aleksander Kolarczyk}
\ead{aleksander.kolarczyk@udose.eu}

\address[SUT]{Silesian University of Technology, Institute of Physics - Centre for Science and Education, Division of Radioisotopes, ul. S. Konarskiego 22B, 44-100 Gliwice, Poland}
\address[AKOTECH]{AKOTECH, ul. Cypriana Norwida 6/10, 41-700 Ruda \'Sl\k{a}ska, Poland}

\begin{abstract}
$\mu$Dose is a novel compact analytical instrument for assessing low level $^{238}$U, $^{235}$U, $^{232}$Th decay chains and $^{40}$K radioactivity. The system is equipped with a dual $\alpha$/$\beta$ scintillator allowing discrimination between $\alpha$ and $\beta$ particles. The unique built-in pulse analyzer measures the amplitude of each individual pulse, its shape and the time interval between subsequent pulses. This allows the detection of pulse pairs arising from subsequent decays of $^{214}$Bi/$^{214}$Po, $^{220}$Rn/$^{216}$Po, $^{212}$Bi/$^{212}$Po and $^{219}$Rn/$^{215}$Po. The obtained $\alpha$ and $\beta$ counts and four separate decay pair counts are used to calculate $^{238}$U, $^{235}$U, $^{232}$Th and $^{40}$K specific activities in measured samples through the use of radioactivity standards.
The $\mu$Dose system may be equipped with various photomultipliers and counting containers to assess radionuclide concentrations of samples of masses ranging between 0.4~g and 4~g. As a result, the user can customize the system to their needs and maximize the instrument's performance. The system is controlled by dedicated software with a graphical user interface and modules for system calibration, data visualization, specific radioactivity calculations and dose rate determination using the infinite matrix assumption.

\end{abstract}


\end{frontmatter}


\section{Introduction}
In trapped charge dating (optically stimulated luminescence, thermoluminescence, electron spin resonance) age is determined from the equivalent total absorbed radiation dose and radiation dose rate. In the natural environment radiation dose rate arises from $^{238}$U, $^{235}$U, $^{232}$Th decay chains and $^{40}$K. Commonly this is assessed using - field instruments, and laboratory analysis of samples including thick source alpha counting, thick source beta counting, high resolution gamma spectrometry, alpha spectrometry, mass spectrometry, flame photometry, inductively couples plasma mass spectrometry, neutron activation, X-ray fluorescence and other techniques. 

Thick source alpha counting (TSAC) has been developed by \cite{Turner1958} to provide a relatively easy and inexpensive way of assessing the low level content of alpha emitters in biological tissues. Later this technique was adapted to determining the uranium and thorium content in samples of fired ceramics for assessing the annual dose for luminescence dating by \cite{Aitken1985}. In TSAC, the powdered sample is placed on a plastic sheet coated with a very thin layer of the scintillator ZnS:Ag attached to its surface. Alpha particles, emitted by U and Th series members, upon reaching the screen produce scintillations with practically 100\% efficiency. In order to estimate the contribution from the U and Th series, additionally the so-called slow and sometimes fast pairs are counted. Such pairs are fast successions of counts due to the short lived $^{216}$Po ($t_{1/2}$~=~0.145~s) in the $^{232}$Th series and $^{215}$Po ($t_{1/2}$~=~1.78~ms) in the $^{235}$U series \citep{Aitken1985}. Today the TSAC technique is widely used \citep{Cawthra2018,Chen2015,Duller2015,Fu2017,Jan2016,Kuhn2017,Sabtu2015,Schmidt2017} for assessing the annual dose in trapped charge dating techniques. However, there are some significant limitations in the conventional TSAC technique. For example, the activity of $^{40}$K which is a major dose contributor to the dose rate in environmental samples, cannot be determined using this technique and an independent determination by different means is usually performed \citep{Dunseth2017,Jacobs2016,Jun2016,Roettig2017}. Additionally, the $^{238}$U decay chain may cause problems due to possible disequilibrium\citep{Eitrheim2016,Krbetschek1994,Prescott1995}. Another drawback is the influence of sample' s reflectance on the TSAC efficiency \citep{Huntley1978} which can cause up to 6\% error in an apparatus setting proposed by \cite{Aitken1985}.\\
\indent In addition to the TSAC technique an alternative method of $\beta$ counting was proposed by \cite{Sanderson1988}. This is a much more rapid technique of annual dose rate determination, however it cannot assess the specific radioactivities of $^{238}$U, $^{235}$U, $^{232}$Th decay chains or $^{40}$K \citep{Sanderson1988,BotterJensen1988}.\\
\indent In the current work, we describe a novel system - $\mu$Dose - that opens up new possibilities and largely removes limitations described in the previous paragraphs. This is done through the employment of a dual $\alpha$/$\beta$ scintillator module, a new measurement setup, a new pulse analyzer unit \citep{Miosz2017}, and advanced data processing. The system allows detection of two $\beta$/$\alpha$ decay pairs in addition to the above mentioned $\alpha$/$\alpha$ pairs. The first pair arises in the $^{232}$Th series from subsequent decays of $^{212}$Bi and $^{212}$Po where $^{212}$Po has a half-life of 299~ns. The second $\beta$/$\alpha$ pair arises in the $^{238}$U series from subsequent decays of $^{214}$Bi and $^{214}$Po where $^{214}$Po has a half-life of 164~$\mu$s. Therefore, four decay pairs can be used to assess the specific activity of thorium and uranium decay chains as well as the potassium activity.\\
\indent The $\mu$Dose system is designed with emphasis on natural radioactivity measurement and the software is equipped with modules for dose rate measurement that is dedicated to trapped charge dating. With some modifications the system may be adapted for other purposes as well.\\
\indent In the following sections we provide a detailed description of the $\mu$Dose system, its performance, decay chain activity measurement method and the $^{40}$K assessment.

\section{Experimental Section}

\subsection{System construction}
\begin{figure}
\includegraphics[width=90mm]{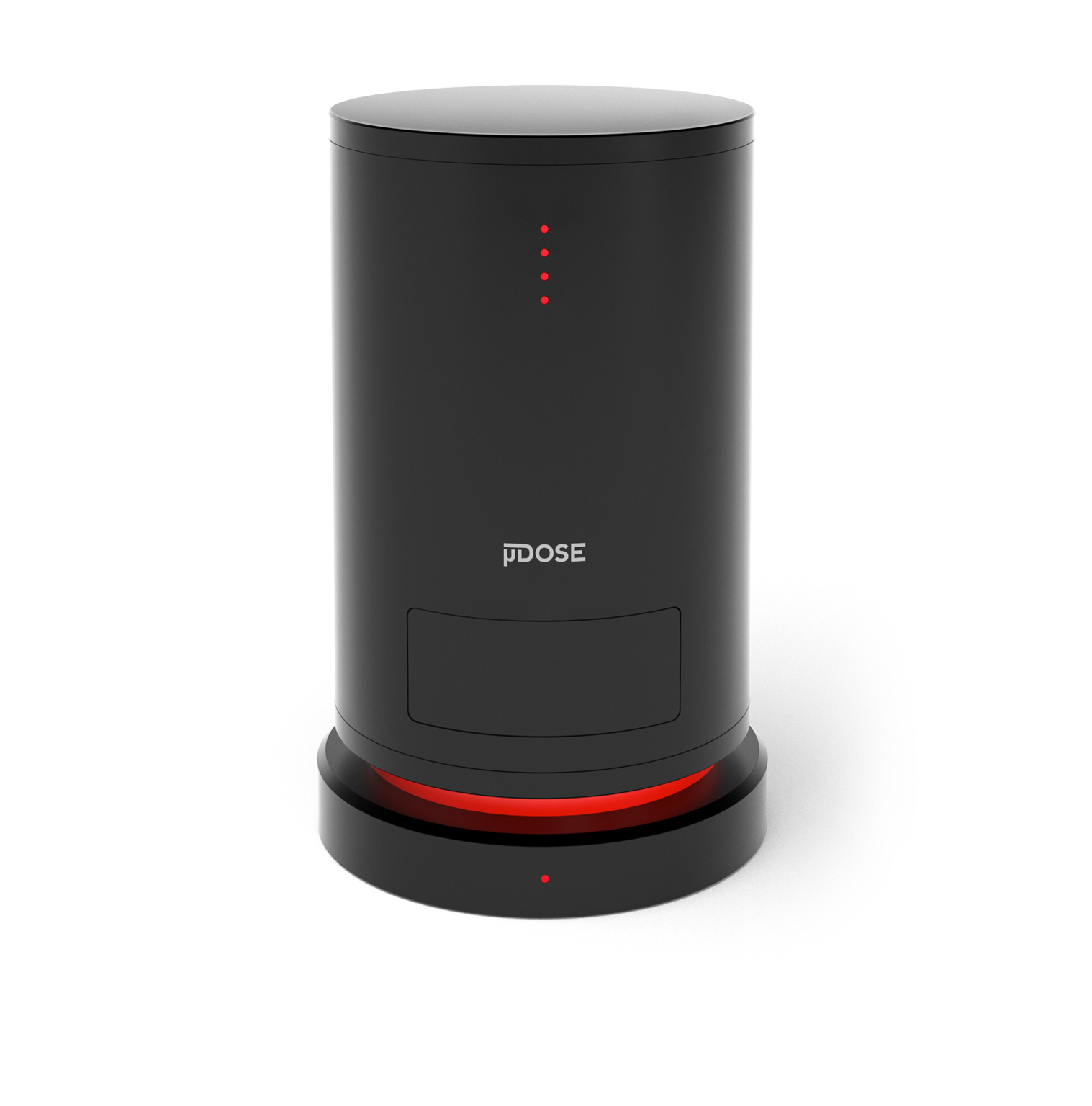}
\caption{\label{Fig1}The $\mu$Dose system.}
\end{figure}
\begin{figure*}
\centering
\includegraphics[width=140mm]{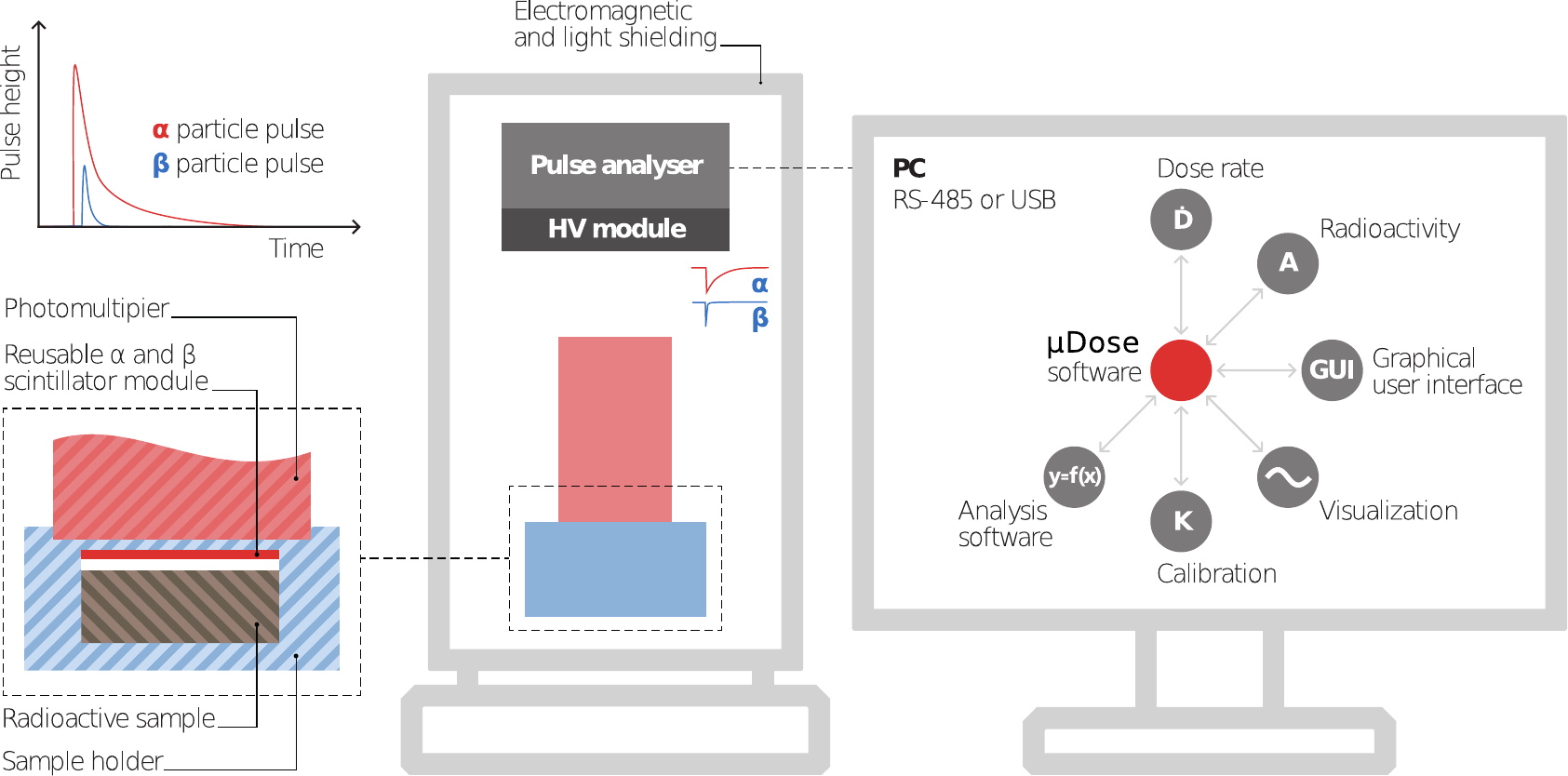}
\caption{\label{Fig2}Block diagram of the $\mu$Dose system.}
\end{figure*}
$\mu$Dose is a very compact system (shown in Fig.~\ref{Fig1}) as it takes just over 20~cm $\times$ 20~cm of desk space and 35~cm height. The entire electronics, including a stable high voltage power supply, a photomultiplier and a pulse analyzer are built into the system and no additional components except a PC (which can control several such devices) are required for system operation.
In the $\mu$Dose system, the sample is placed in a dedicated sample container which is placed below the photomultiplier (Fig.~\ref{Fig2}). The container is equipped with a reusable dual (sandwich)  $\alpha$ and $\beta$ scintillator module covered by a 0.2~$\mu$m replaceable silver foil on the underside of the scintillator. The silver foil is easily penetrated by $\beta$ and the vast majority of $\alpha$ particles emitted by the sample. The silver layer increases the number of photons that reach the photomultiplier (PMT) and it also removes the measurement variability that arises from the sample's reflectance because scintillation photons are reflected from the silver layer rather than the sample. The sample container is gas-tight to prevent radon migration from and into it. The sample itself has a geometry of a thin disk which matches the diameter of the photocathode. Depending on the expected sample mass, the system can accommodate PMT's which have photocathode diameter from 30~up to 70~mm.

\subsection{Electronics}
The pulse analyzer has been described in detail in \cite{Miosz2017}, therefore here only a brief description is given. The $\alpha$ and $\beta$ particles produce scintillations in two different scintillator layers. The generated pulse shapes are different for each  of the two scintillators, permitting the identification of the source particle of each pulse. This shape is preserved by the PMT and amplifier where the scintillations are transformed into electrical pulses and significantly amplified. The pulse analyzer detects the incoming voltage pulses (Fig.~\ref{Fig2}) and stores them as series of ADC values that represent each pulse. These pulse data are time-stamped and stored for further processing. The acquired data are then transferred to the computer and processed by a dedicated algorithm that determines the pulse height, the pulse shape and the time when each pulse appeared. The pulse height and the pulse shape allow to discriminate between $\alpha$ and $\beta$ induced pulses. In addition, the algorithm is also capable of identifying pulses that do not match neither $\alpha$ nor $\beta$ particles \citep{Tudyka2017b} enabling the removal of background pulses arising from electrical noise or other interfering sources. The software also deconvolutes piled up pulses from decays that appeared within a small time interval (ca. 100~ns).\\
\indent The electronics module has a built-in high stability, low ripple, high voltage supply. This is controlled and monitored for system stability by the software. To protect the PMT, the high voltage is automatically switched off when the drawer with the sample container is opened.\\
\begin{figure}
\centering
\includegraphics[width=75mm]{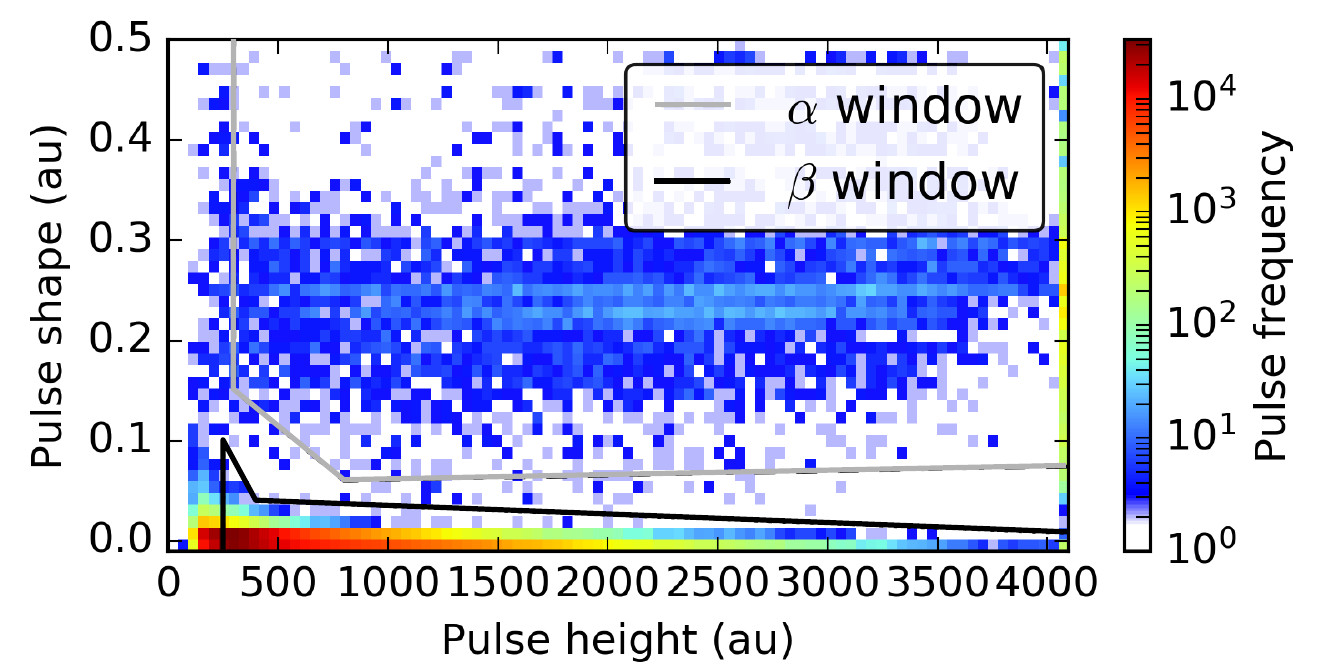}
\caption{\label{Fig3}Data for a sample no. 1 (see section~\ref{section1}) measured for 44~hours. 2D pulse height vs. pulse histogram where the colour indicates relative frequency of appearance of the pulses. The $\alpha$ window is marked with the grey line above ca. 0.1 au pulse shapes parameter, the $\beta$ window is marked with the black line below the $\alpha$ window. Note that beyond the $\alpha$ window there are pulses whose height exceeds 4000 au.}
\end{figure}

\begin{figure*}
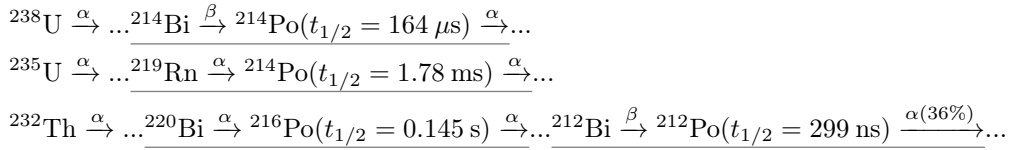

\begin{equation*}
\begin{aligned}
& \text{\textsuperscript{238}U}  \xrightarrow[]{\alpha} ... \Cline[gray]{\text{\textsuperscript{214}Bi} \xrightarrow[]{\beta} \text{\textsuperscript{214}Po} (t_{1/2}=164 \: \mu \text{s}) \xrightarrow[]{\alpha}} ...\\ 
& \text{\textsuperscript{235}U}   \xrightarrow[]{\alpha} ... \Cline[gray]{^{219}\text{Rn} \xrightarrow[]{\alpha} \text{\textsuperscript{214}Po}(t_{1/2}=1.78 \: \text{ms}) \xrightarrow[]{\alpha}} ...\\
& \text{\textsuperscript{232}Th}   \xrightarrow[]{\alpha} ... \Cline[gray]{^{220}\text{Bi} \xrightarrow[]{\alpha} \text{\textsuperscript{216}Po} (t_{1/2}=0.145 \: \text{s}) \xrightarrow[]{\alpha}} ...  \Cline[gray]{\text{\textsuperscript{212}Bi} \xrightarrow[]{\beta} \text{\textsuperscript{212}Po} (t_{1/2}=299 \:\text{ns}) \xrightarrow[]{\alpha(36\%)}} ...
\end{aligned}
\end{equation*}
\caption{\label{decay_chains}The sections of $^{238}$U, $^{235}$U and $^{232}$Th decay chains where the decay pairs occur showing relevant decay modes, half-lifes and the branching ratio where relevant. }
\end{figure*}

\subsection{$\alpha$, $\beta$ and decay pairs detection}

Fig.~\ref{Fig3} shows a typical 2D pulse height vs. pulse shape histogram where the colour indicates relative frequency of the recorded pulses obtained from sample 1 with artificial $^{238}$U, $^{235}$U, $^{232}$Th decay chains and $^{40}$K concentrations (see paragraph ''Samples and sample preparation'' for detailed description).

\begin{figure*}[!htb]
\centering
\includegraphics[width=140mm]{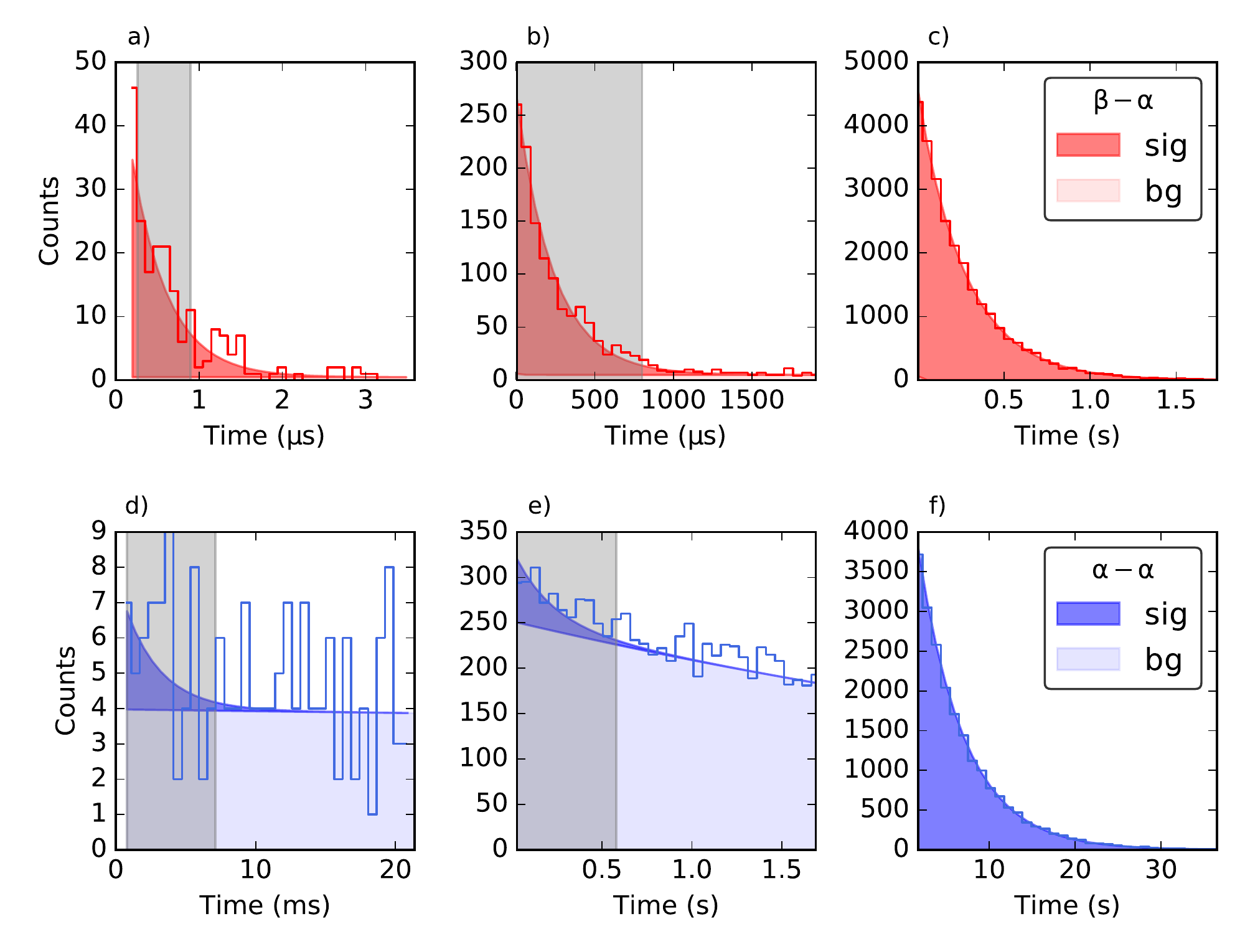}
\caption{\label{Fig4}Data for sample no. 1 measured for 44~hours. Distributions for $\beta$-$\alpha$ time intervals in ranges a) 0-3.6~$\mu$s, b) 3.6-2100~$\mu$s, c) >~2100~$\mu$s and $\alpha$-$\alpha$ time intervals in ranges d) 0.5-21~ms, e) 21~ms - 1.7~s, c) >~1.7~s. Measured time intervals are shown by the stepped lines, the fitted signal (s) and background (b) are depicted by the filled areas. a), b), d) and e) plots reveal subsequent decays of $^{220}$Rn/$^{216}$Po, $^{219}$Rn/$^{215}$Po, $^{212}$Bi/$^{212}$Po and $^{214}$Bi/$^{214}$Po, respectively. d) and g) show remaining, purely random $\beta$-$\alpha$ and $\alpha$-$\alpha$ time interval distributions. Note the differing horizontal axis scales on plots a-f.}
\end{figure*}
In environmental samples, of typically 100-1000~Bq$\cdot$kg$^{-1}$ total activity concentration, uncorrelated decays would be detected at a rate of <~1~s$^{-1}$. However, as can be seen in diagram (Fig.~\ref{decay_chains}), some radionuclides in the natural decay chains have half-lifes significantly less than 1~s, which would result in correlated decays with much shorter time intervals observed as two pulses being detected in quick succession, so-called decay pairs. Since $\mu$Dose detects and identifies both $\alpha$ and $\beta$ particles, the system can identify and count the two $\alpha$-$\alpha$ decay pairs used in TSAC ($^{220}$Rn/$^{216}$Po, $^{219}$Rn/$^{215}$Po) and also two $\beta$-$\alpha$ decay pairs ($^{212}$Bi/$^{212}$Po, $^{214}$Bi/$^{214}$Po). To detect those decay pairs the system builds a time interval distribution from the recorded time intervals between events classified as $\beta$ pulses followed in quick succession of an $\alpha$ pulse (Fig.~\ref{Fig4}a, b and c). This reveals the exponential function arising from $^{212}$Bi/$^{212}$Po (Fig.~\ref{Fig4}a) and $^{214}$Bi/$^{214}$Po (Fig.~\ref{Fig4}b) decay pairs. Similarly, time interval distribution between events classified as $\alpha$ (Fig.~\ref{Fig4}d, e and f) allow to reveal the exponential decay arising from $^{219}$Rn/$^{215}$Po (Fig.~\ref{Fig4}d) and $^{220}$Rn/$^{216}$Po (Fig.~\ref{Fig4}e) decay pairs. Fig.~\ref{Fig4}c and f show remaining $\beta$-$\alpha$ and $\alpha$-$\alpha$ time intervals arising from random decays. On each inset in Fig.~\ref{Fig4}, the fitted signal (sig) and background (bg) are marked. Note that background in Fig.~\ref{Fig4}a, b, c, f is virtually absent.

\subsection{$^{238}$U, $^{235}$U, $^{232}$Th and $^{40}$K activity assessment}
The four decay pairs count rates can be used to directly determine the activity per unit of mass 
\begin{equation}
\begin{aligned}
& r_{Bi-212/Po-212}=k_{Bi-212/Po-212}a_{Bi-212/Po-212},\\\label{eq1}
& r_{Bi-214/Po-214}=k_{Bi-214/Po-214}a_{Bi-214/Po-214},\\
& r_{Rn-220/Po-216}=k_{Rn-220/Po-216}a_{Rn-220/Po-216},\\
& r_{Rn-219/Po-215}=k_{Rn-219/Po-215}a_{Rn-219/Po-215}.\\ 
\end{aligned}
\end{equation}
Here $r$ are the net count rates of decay pairs indicated in subscripts, $k$ are calibration parameters for the given system and the given decay pairs indicated in subscripts, and $a$ are specific activities of decay pairs indicated in subscripts. The $r$ net count rates are obtained from the total and background pair events. Eqs. (\ref{eq1}) hold for samples with atomic compositions similar to those of the calibration standards used. Eqs. (\ref{eq1}) remain valid regardless of the state of secular equilibrium in the measured material.\\
\indent In many cases, samples are close to secular equilibrium and the user may assume the following relationships 
\begin{equation}
\begin{aligned}
& a_{Th-232} = a_{Bi-212/Po-212}=a_{Rn-220/Po-216},\\\label{eq2}
& a_{U-238} = a_{Bi-214/Po-214},\\
& a_{U-235} = a_{Rn-219/Po-215},\\
& r_{\alpha} = k_{\alpha, Th-232}a_{Th-232}+k_{\alpha, U-238}a_{U-238}\\&\;\;\;\;\;\;\;+k_{\alpha, U-235}a_{U-235},\\
& r_{\beta} = k_{\beta, Th-232}a_{Th-232}+k_{\beta, U-238}a_{U-238}\\&\;\;\;\;\;\;\;+k_{\beta, U-235}a_{U-235}+k_{\beta, K-40}a_{K-40}.
\end{aligned}
\end{equation}

Here $r_\alpha$ and $r_\beta$ are the $\alpha$ and $\beta$ net count rates. 
Eqs. (\ref{eq1}) and (\ref{eq2}) enable the calculation of the decay chain specific activities in the sample. These equations calculate the activity of pure $\beta$ emitters. In (Eg. \ref{eq2}), it is assumed that $^{40}$K is the major $\beta$ contributor, which is true for most environmental samples. However, it needs to be borne in mind that in environmental samples also other $\beta$ emitters can be found, e.g. natural $^{87}$Rb or anthropogenic $^{137}$Cs \citep{Ochiai2018,Buesseler2015,por2015,Faure1972,Evangeliou2014,Sanderson2016}. This needs to be considered individually for each sample as it may introduce an error in the $^{40}$K assessment. For example \cite{Warren1978} gives an average $^{87}$Rb at the level of 50 ppm of natural rubidium per 1\% of potassium. \\
\indent Eqs. (\ref{eq1}) and (\ref{eq2}) can be further restricted using a known $^{238}$U/$^{235}$U isotopic ratio \citep{Uvarova2014,Brennecka2010}
\begin{equation}
\frac{^{238}U}{^{235}U}=\frac{a_{U-238}/\lambda_{U-238}}{a_{U-235}/\lambda_{U-235}}=137.88.\label{eq3}
\end{equation}
where $\lambda$ is decay constant of radioisotope indicated in subscript. Eq. (\ref{eq3}) removes one degree of freedom and allows more precise results to be obtained. In the natural environment, the ratio $^{238}$U/$^{235}$U ratio can vary by up to ca. 5\textperthousand \citep{Uvarova2014,Brennecka2010,Placzek2016,Phan2018,Brennecka2018}. This however is consistent enough to assume it is constant within measurement accuracy and precision. 
\subsection{System calibration}
The $\mu$Dose system needs to be calibrated with reference materials of known radioactivities, as well as a background sample. In the current work we use IAEA-RGU-1, IAEA-RGTh-1, and IAEA-RGK-1 standards from the International Atomic Energy Agency  \citep{IAEA}. The IAEA-RGU-1 and IAEA-RGTh-1 are produced using uranium and thorium ores that are mixed with floated silica powder. Decay chains present in those reference materials can be considered to be in secular equilibrium with parent radioactivity. The IAEA-RGK-1 reference material is produced using high purity (99.8\%) potassium sulfate. 
The $\mu$Dose software contains a dedicated module that allows the user to easily obtain calibration parameters ($k$ in Eqs. \ref{eq1} and \ref{eq2}) by means of weighted least squares method. This is done by matching calibration measurements with known radioactivities from the built-in database. 
\section{Results and Discussion}

\subsection{Samples and sample preparation}\label{section1}

To test the performance of $\mu$Dose, activities of five samples were assessed using two additional systems, namely, a high-purity germanium (HPGe) $\gamma$ spectrometer and a conventional TSAC system.\\
\indent  Sample 1 was an artificial sample composed from IAEA-RGU-1, IAEA-RGTh-1, and IAEA-RGK-1 mixed in equal weight proportions allowing the calculation of its specific activities using the IAEA reference activities.  Samples 2, 3 and 4 are loess sediments. Sample 5 is a brick from archaeological excavations.\\
\indent  For $\mu$Dose and TSAC, the samples were powdered using an agate mill to avoid overcounting of $\alpha$ particles which may arise from inhomogeneous distribution of radioactive elements in natural samples \citep{Zoller1989,Murray1982,Poreba2006}. After the milling grain size distribution was verified using a laser diffractometer Mastersizer 3000 manufactured by Malvern Instruments Ltd. For all samples the median particle size was less than 7~$\mu$m. In both $\mu$Dose and TSAC systems, we used ground up 1~g samples and 42~mm diameter scintillators. The measurement time was the same for the $\mu$Dose and TSAC systems.
TSAC measurements were performed using an in-house built system with pulse amplitude-time analyzer \cite{Tudyka2011}. 
\indent The $\gamma$ spectrometry measurements were performed using a low background, high resolution HPGe detector with a resolution at full width at half maximum of 1.8~keV and relative efficiency of 40\% at the energy of 1332~keV manufactured by Canberra. The same standards were used as above.  The average specific radioactivities were calculated using a weighted mean obtained for selected lines. The lines $^{234}$Th, $^{234m}$Pa, $^{214}$Pb, $^{214}$Bi and $^{210}$Pb were used for  $^{238}$U, whereas the lines of $^{228}$Ac, $^{212}$Pb, $^{212}$Bi and $^{208}$Tl were used for $^{232}$Th.\\

\subsection{Measurements}

The system set-up and data were evaluated according to \cite{Aitken1985}.
$\mu$Dose specific uranium, thorium and potassium radioactivities were obtained using Eqs. (\ref{eq1}-\ref{eq3}).\\

\indent The results and counting times are summarised in Table~\ref{tab1}.

\begin{table*}
\caption{Specific radioactivity measurements using $\mu$Dose, a HPGe and a traditional TSAC system. Given uncertainties correspond to $1 \sigma$.}
\centering
\begin{adjustbox}{width=1\textwidth}
  
  \label{tab1}
  \begin{tabular}{ccccccccccc}
    \hline
    & \multicolumn{3}{c}{Mean $^{238}$U radioactivity}  & \multicolumn{3}{c}{Mean $^{232}$Th radioactivity} & \multicolumn{2}{c}{$^{40}$K assessment} & \multicolumn{2}{c}{Measurement time} \\
    & \multicolumn{3}{c}{(Bq$\cdot$kg$^{-1}$)}  & \multicolumn{3}{c}{(Bq$\cdot$kg$^{-1}$)} & \multicolumn{2}{c}{(Bq$\cdot$kg$^{-1}$)} & \multicolumn{2}{c}{(hr)} \\
    \hline
    & $\mu$Dose  & TSAC & HPGe & $\mu$Dose  & TSAC & HPGe & $\mu$Dose  & HPGe & $\mu$Dose \& TSAC & HPGe \\
    
    1\textsuperscript{\emph{a}} & 1620 $\pm$ 40 & 2400 $\pm$ 150 & 1628 $\pm$ 32 & 1110 $\pm$ 60 &  1300 $\pm$ 150 & 1062 $\pm$ 37 & 4480 $\pm$ 160 & 4610 $\pm$ 110 & 44 & 6.5 \\
    2 & 26.3 $\pm$ 2.6 & 30.1 $\pm$ 4.4 & 26.08 $\pm$ 0.59 & 32 $\pm$ 4.0 & 31.8 $\pm$ 4.3 & 33.90$\pm$ 1.10 & 576 $\pm$ 48 & 564 $\pm$ 19 & 74 & 25 \\
    3 & 30.9 $\pm$ 2.5 & 26.7 $\pm$ 4.2 & 26.16 $\pm$ 0.55 & 30.1 $\pm$ 3.3 & 39.3 $\pm$ 4.1 & 32.86$\pm$ 1.05
 & 588 $\pm$ 20 & 532 $\pm$ 18 & 98 & 27 \\
    4 & 38.1 $\pm$ 3.9 & 41.6 $\pm$ 5.2 & 27.71 $\pm$ 0.58 & 33.6 $\pm$ 4.5 & 41.1 $\pm$ 5.2 & 35.46 $\pm$ 1.13 & 618 $\pm$ 27 & 570 $\pm$ 19 & 66 & 24 \\
    5 & 23.5 $\pm$ 2.8& 26.4 $\pm$ 2.4 & 17.80 $\pm$ 0.37 & 19.3 $\pm$ 2.8 & 24.0 $\pm$  2.4 & 19.48 $\pm$ 0.89 & 308 $\pm$ 18 & 324 $\pm$ 11 & 87 & 29 \\
    \hline
  \end{tabular}
\end{adjustbox}
\textsuperscript{\emph{a}}  Sample created form mixing IAEA-RGU-1, IAEA-RGTh-1, and IAEA-RGK-1 in equal weight proportions. The activities calculated using reference values\citep{IAEA} are 1673~Bq$\cdot$kg$^{-1}$ of $^{238}$U, 1083~Bq$\cdot$kg$^{-1}$ of $^{232}$Th and 4669~Bq$\cdot$kg$^{-1}$ of $^{40}$K.
\end{table*}

\subsection{System performance}

As seen in Table~\ref{tab1}, there is a very good agreement between the values obtained using the $\mu$Dose, gamma spectrometry and reference value for sample 1 (a mix of the IAEA standards - see previous subsection). In this case, the TSAC result significantly deviates from the known activities. This might be caused by a different sample reflection and a radioactivity much higher than seen in average sediment samples. For all samples, the results obtained using TSAC are characterised by larger measurement errors for the same counting times, and as mentioned earlier information on the potassium content is unavailable. 

Thorium and potassium specific activities agree within 2 standard deviations between $\mu$Dose and gamma spectrometry. Uranium specific activities for samples 4 and 5 agree within 3 standard deviations between $\mu$Dose and gamma spectrometry. In the investigated samples, there was no indication of possible lack of secular equilibrium in the U and Th decay chains. 

When the activities are used to estimate the annual dose in trapped charge dating applications it has to be borne in mind that the values and errors returned by $\mu$Dose are correlated. This fact is taken into account during the calculation of the annual dose and leads to lower dose rate errors than in case these values were independent , as e.g. in high resolution gamma spectrometry. The calculations of annual dose will be discussed elsewhere.

\section{Conclusion}
The $\mu$Dose system allows to detect $\alpha$ and $\beta$ radiation with four different decay pairs arising in the $^{238}$U decay chain ($^{214}$Bi/$^{214}$Po), $^{232}$Th decay chain ($^{220}$Rn/$^{216}$Po and $^{212}$Bi/$^{212}$Po) and $^{235}$U decay chain ($^{219}$Rn/$^{215}$Po). If the sample is close enough to secular equilibrium, the obtained $\alpha$ and $\beta$ counts and four separate decay pairs allow to obtain the $^{238}$U, $^{235}$U and $^{232}$Th decay chains concentration in the sample. The $^{40}$K activity is assessed from the excess of $\beta$ counts over what is expected over what is predicted from $^{238}$U, $^{235}$U and $^{232}$Th measurements.

$\mu$Dose software allows for a convenient system calibration which limits routine work required from the user. The activities are calculated according to various assumptions on secular equilibrium, and measurement reports are automatically created for convenient post-processing.
The $\mu$Dose system can be equipped with various photomultipliers to assess various sample masses from 0.4--4~g. The software is build with emphasis on EPR/OSL/TL dating therefore it includes various modules for dose rate calculation.

\section{Acknowledgements}

The development of the pulse analyzer used in the  $\mu$Dose system  was supported with the grant LIDER/001/404/L-4/2013 by the Polish National Centre for Research and Development.
Currently the project is co-financed by the Polish Ministry of Science and Higher Education from ''Incubator of Innovation+'' programme within the framework of the Smart Growth Operational Programme, Action 4.4 Potential increase of human resources of the R\&D sector.

The authors thank Ms. Agnieszka Szymak for her help in carrying out the measurements. 

\section{References}

\bibliography{uDose}

\end{document}